\documentclass[9pt,twocolumn,twoside]{osajnl}

\journal{ol} 

\setboolean{shortarticle}{true} 

\title{Bulk-edge correspondence and trapping at a non-Hermitian topological interface}
\author[1,2,*]{Stefano Longhi}
\affil[1]{Dipartimento di Fisica, Politecnico di Milano, Piazza L. da Vinci 32, I-20133 Milano, Italy}
\affil[2]{IFISC (UIB-CSIC), Instituto de Fisica Interdisciplinar y Sistemas Complejos - E-07122 Palma de Mallorca, Spain}
\affil[*]{Corresponding author: stefano.longhi@polimi.it}

\dates{Compiled \today}

\doi{\url{http://dx.doi.org/10.1364/ol.XX.XXXXXX}}
\begin{abstract}
In Hermitian systems, according to the bulk-edge correspondence interfacing two topological optical media with different bulk topological numbers implies the existence of  edge states, which can trap light at the interface. However, such a general scenario can be violated when dealing with non-Hermitian systems. Here we show that interfacing two semi-infinite Hatano-Nelson chains with different bulk topological numbers can result in the existence of infinitely many edge (interface) states, however light waves cannot be rather generally trapped at the interface. 
 \end{abstract}
 
\setboolean{displaycopyright}{true}

\begin{document}

\maketitle
\thispagestyle{fancy}

{\em Introduction.} A central result of topological materials is the bulk-edge correspondence:  when two materials with different bulk topological invariants are interfaced, there should exist edge
states spatially localized at the interface and with energies that lie within the energy gap of the surrounding bulk media \cite{r1,r2,r3}. Any initial excitation spatially localized near the interface populates rather generally propagating (extended) states, that spread in the bulk, and edge states that remain trapped at the interface. As a result, at least a fraction of the initial excitation remains trapped at the interface for ever. However, such a trapping scenario can be deeply modified when turning to non-Hermitian (NH) systems, where spectral and dynamical localization become rather distinct concepts \cite{r4}.
The recent discovery of topological properties of NH systems \cite{r5,r6,r7,r8} underpins major phenomena such as the NH skin effect, i.e. the macroscopic condensation of bulk eigenstates at the edges under open boundary conditions (OBC), and a generalized bulk-edge correspondence \cite{r5,r6,r7,r8,r9,r10,r11,r12,r13,r14,r15,r16,r17,r18,r19,r20,r21,r22,r23,r24,r25,r26}. The NH skin effect is observed rather generally when
a synthetic imaginary gauge field is applied to an otherwise Hermitian  system. A paradigmatic example is provided by the one-dimensional tight-binding lattice with an imaginary gauge field $h$, originally introduced by Hatano and Nelson \cite{r27}. The asymmetric coupling between adjacent sites in the lattice yields the NH skin effect under OBC, and a robust biased transport in the bulk \cite{r28,r29}, which is reversed when the sign of $h$ is flipped.  Since the energy spectrum of the Hatano-Nelson model under periodic boundary conditions (PBC) is complex and forms a closed loop in the complex plane [Fig.1(a)], it is characterized by a non-zero winding number $w= \pm1$ for any base energy $E_B$ inside the closed loop \cite{r5}. This result is in sharp contrast with the general theory of Hermitian topological insulators in one dimension, where a non-trivial topology requires at least two bands and chiral symmetry. Under semi-infinite boundary conditions (SIBC), a bulk-edge correspondence can be established \cite{r5,r18}: an edge state, localized either at the left or right boundaries (depending on the sign of $h$), does exist for any complex energy $E_B$ inside the closed loop.\\ 
In this Letter we consider a topological interface, obtained by connecting two semi-infinite Hatano-Nelson chains with different values of the gauge field $h$, as shown Figs.1(b) and (c). When the gauge fields are of opposite sign and  the biased flow in the bulk of the two media is directed toward the interface,  a topological funnel effect is observed \cite{r19} (see also \cite{r14}),  associated to the existence of  infinitely many localized interface states: for any arbitrary excitation of the lattice, light is directed toward the interface and remains  there trapped. However, a different behavior is found when the gauge fields in the two regions have the same sign, so as light is pushed toward the interface from one side but pulled outward the interface from the other side. Naively, one could think that in this case light can never be trapped at the interface and that interface states should not exist, because light is pulled outward the interface from one side. Contrary to such an intuitive picture, we show here that there are infinitely many exponentially localized interface states at any complex energy corresponding to different topological numbers in the bulk of the two media, thus establishing a bulk-edge correspondence for the topological interface. However, unlike the previous case light can be or cannot be trapped at the interface, depending on the initial excitation condition. In particular, for any initial spatially-localized excitation of the lattice with a faster-than-exponential localization, light cannot be trapped at the interface.\\	

{\em Topological interface and the bulk-edge correspondence.} The Hatano-Nelson model with an inhomogeneous imaginary gauge field $h=h(n)$ is described in physical space by coupled equations \cite{r5,r18,r27,r29}
\begin{equation}
i \frac{d \psi_n}{dt}= \Delta \left\{ \exp [ h(n+1)] \psi_{n+1}+ \exp [-h(n)] \psi_{n-1} \right\}
\end{equation}
for the wave amplitudes $\psi_n$ at the various lattice sites, where $\Delta \exp(\pm h)$ are the left/right hopping amplitudes. Let us first briefly recall the topological properties of the model in the homogeneous case $h(n)=h \; {\rm constant}$ \cite{r5}. Under PBC, with the Ansatz $\psi_n=\exp(ikn)$ the Hamiltonian in Bloch space  reads $H(k)=2 \Delta \cosh(h+ik)$, where $-\pi \leq k < \pi$ is the Bloch wave number. The corresponding energy spectrum describes a closed loop (an ellipse) in complex plane[Fig.1(a)]. For a given complex base energy $E_B$, a winding number $w(E_B)$ can be introduced \cite{r5,r18}
\begin{equation}
w(E_B)=\frac{1}{2 \pi i} \int_{-\pi}^{\pi} dk \log \left\{ H(k)-E_B \right\}.
\end{equation}
Clearly, one has $w(E_B)=0$ when $E_B$ is external to the ellipse, while $w(E_B)=h/ |h|= \pm 1$ when $E_B$ is internal to the ellipse. Under OBC, the energy spectrum collapses to the segment $(-\Delta,\Delta)$ on the real axis and bulk states become squeezed toward one of the edges (skin effect), while for SIBC the energy spectrum is the interior of the ellipse \cite{r5,r18}. In the bulk, a backward (forward) biased drift along the lattice is observed for $h>0$ ($h<0$), with a velocity $v=2 \Delta \sinh |h|$ \cite{r12}.\\ 
Let us then turn to the inhomogeneous case, specifically we consider an interface with $h(n)=h_1$ for $n \leq 0$ and $h(n)=h_2$ for $n>0$, as shown in Figs.1(b) and (c). Note that, after the non-unitary gauge transformation
\begin{equation}
\psi_n=c_n \exp[-V(n)]
\end{equation} 
with $V(n)=\sum_{l=1}^n h(l)$ for $n \geq 1$, $V(n)=-\sum_{l=0}^{n+1} h(l)$ for $n<0$ and $V(0)=0$, Eq.(1) is formally reduced to the Hermitian tight-binding lattice model with uniform hopping amplitude, i.e.
\begin{equation}
i \frac{dc_n}{dt}=\Delta (c_{n+1}+ c_{n-1}).
\end{equation}
For a stepwise $h(n)$, describing an interface, one has 	
\begin{equation}
V(n)=
\left\{
\begin{array}{ll}
nh_2 &  n \geq 1 \\
nh_1 & n \leq 0.
\end{array}
\right.
\end{equation}
 The formal eigenfunctions to Eq.(4) are given by $c_n=\exp(ikn+\mu n-iEt)$ with energy $E=2 \Delta \cosh(\mu+ik)$, where $k$ and $\mu$ are arbitrary real numbers. From Eqs.(3) and (5), it follows that 
 \begin{equation}
 \psi_n= \exp[ikn+\mu n -iEt-V(n)]
 \end{equation}
  \begin{figure}[htb]
\centerline{\includegraphics[width=8.7cm]{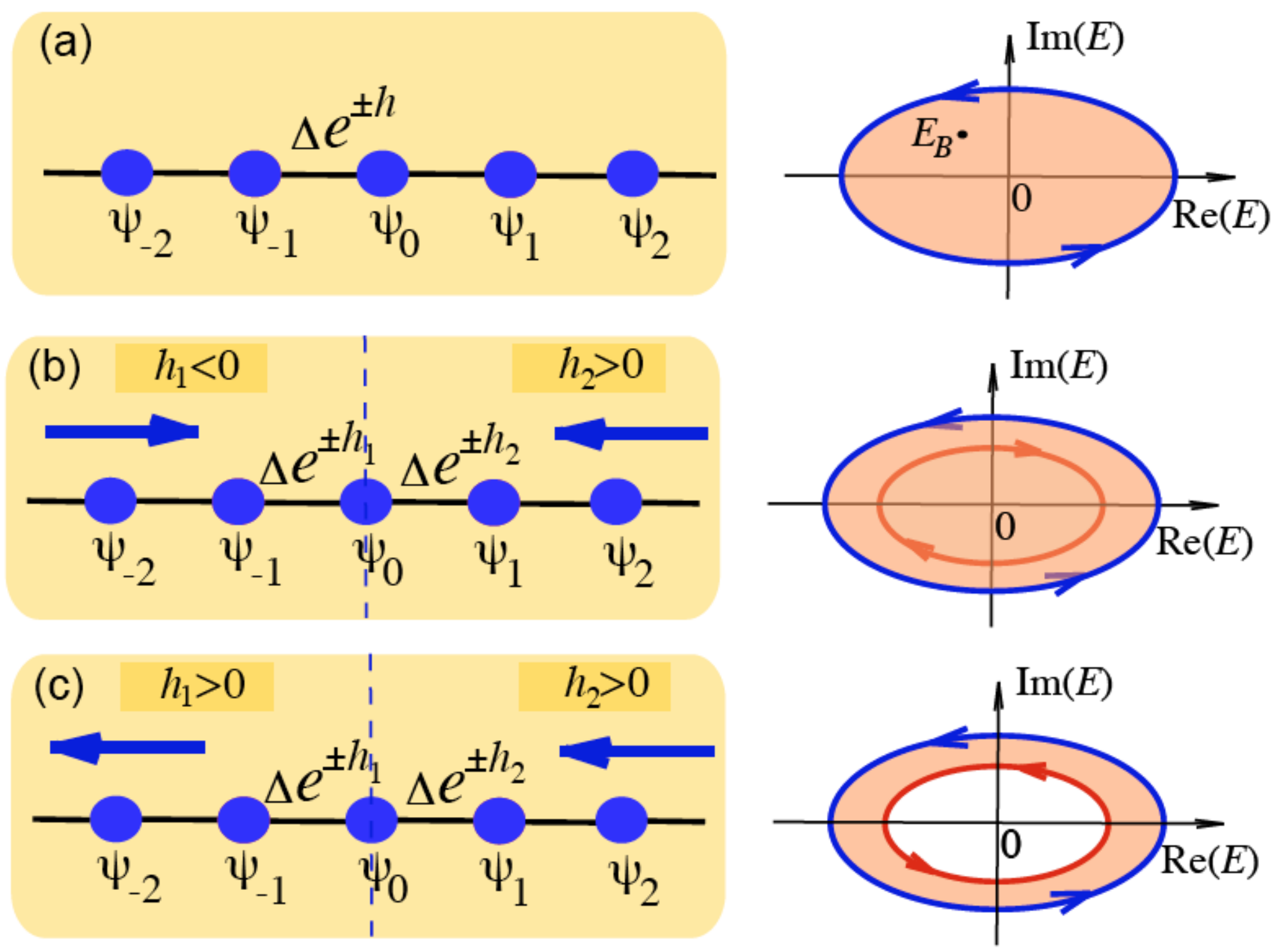}} \caption{ \small
(Color online) Bulk-edge correspondence in a NH interface. (a) Schematic of the Hatano-Nelson lattice model with asymmetric hopping amplitudes $\Delta \exp ( \pm h)$. The PBC energy spectrum is an ellipse in complex energy plane (right panel), which is traveled clockwise for $h<0$ and counter-clockwise for $h>0$. For an energy $E_B$ in the interior of the ellipse, the winding number $w(E_B)$ is 1 for $h>0$ and $-1$ for $h<0$, while for $E_B$ in the exterior of the ellipse one has $w(E_B)=0$. According to the bulk-edge correspondence, the interior of the ellipse (shaded area in the figure) corresponds to the energies of edge states under SIBC. (b) An interface of two Hatano-Nelson chains with imaginary gauge fields $h_1$ and $h_2$, with $h_1<0$ and $h_2>0$. The bold arrows in the left panel indicate the biased flow of excitation in the bulk of the two chains. The right panel shows the two ellipses, corresponding to the PBC energy spectra of the two bulk lattices, which are traveled in opposite directions. According to the bulk-edge correspondence, interface states do exist for any complex energy $E$ in the interior of the outer ellipse (shaded area in the figure), where $w_1(E) \neq w_2(E)$. (c) Same as (b), but for $0 \leq h_1 <h_2$. According to the bulk-edge correspondence, in this case interface states do exist for any complex energy $E$ in the corona between the outer and inner ellipse (shaded area in the figure), where $w_1(E) \neq w_2(E)$.}
\end{figure} 
   remains limited as $n \rightarrow \pm \infty$ provided that $ h_1 \leq\mu \leq h_2$, and $\psi_n$ is exponentially localized around $n=0$ when strictly $ h_1 < \mu < h_2$. The growth (decay) rate $g$ of the eigenstate $\psi_n$ is given by the imaginary part of the energy $E$ and reads
 \begin{equation}
g \equiv {\rm Im}(E)=2 \Delta \sinh \mu \sin k. 
  \end{equation}
    Therefore, edge (interface) states with exponential localization do exist provided that $h_1<h_2$. We are thus facing with two main cases: (i) $h_1<0$, $h_2>0$ and (ii) $0 \leq h_1 <h_2$, which are illustrated in Fig.1(b) and (c) [note that the case $h_1<h_2 \leq 0$ is equivalent to (ii)]. In the former case (i) the energies $E$ of the interface states fill the interior of the outer ellipse, describing the PBC energy spectrum  of the bulk medium with the higher value of $|h|$[see Fig.1(b)]. In the latter case (ii)  the energies $E$ of the interface states fill the corona whose outside and inside boundaries are two ellipses describing the PBC energy spectra of the two interfaced bulk media. Such a result provides the bulk-edge correspondence for the topological interface, since an interface state does exist for any complex energy $E$ such that the corresponding winding numbers $w_1$ and $w_2$ in the two bulk media are different, i.e. for which $w_1(E) \neq w_2(E)$.\\
\\ 
{\it Light trapping at the interface.}  As interface states do exist in both cases of Figs.1(b) and (c) according to the bulk-edge correspondence, light trapping at the interface behaves very different in the two cases. Intuitively, in the former case [Fig.1(b)] light is pushed toward the interface from both sides, and hence one expects light localization at the interface, without any radiating wave, for an arbitrary initial excitation of the lattice. In this regime the interface behaves like a topological funnel for light \cite{r19}, albeit its undergoes an irreversible asymptotic decay in time. Conversely, in the latter case [Fig.1(c)] light is pushed toward the interface from one side, but it is pulled outward the interface from the other side at a slower rate (this follows for the condition $h_1<h_2$). While in this case the bulk-edge correspondence still ensures the existence of infinitely-many localized interface states, an open question is whether any light wave can be trapped by the interface. To this aim, let us first notice that the exact solution to Eq.(1) for an arbitrary initial spatially-localized excitation $\psi_n(0)$ of the lattice can be given in terms of Bessel functions and reads
\begin{equation}
\psi_n(t)=\sum_l \psi_l(0) J_{n-l}( 2 \Delta t) \exp \left[ i \frac{\pi}{2}(l-n) +V(l)-V(n) \right]. 
\end{equation}
Such a relation follows straightforwardly from the well-known impulse response of Eq.(4) and the non-unitary gauge transformation (3). However, Eq.(8) is not much useful to establish whether light trapping can be observed. A different approach, based on an asymptotic analysis, is illustrated in the Supplemental document. For $0 \leq h_1 < h_2$, it can be shown that: \\
{\em Theorem I}. For any initial excitation of the lattice $\psi_n(0)$, spatially-localized with a higher-than-exponential localization, light trapping is never observed.\\
{\em Theorem II}. An interface state with complex energy $E$ is stable against compact deformations provided that $g={\rm Im}(E)$ is larger than $g_0 \equiv 2 \Delta \sinh h_1$.\par
The above results are illustrated in Figs.2, 3 and 4. The figures show the time evolution of the normalized amplitudes $a_n(t)=\psi_n(t) / \sqrt{P(t)}$, where $P(t)=\sum_n |\psi_n(t)|^2$ is the total optical light power, for different initial excitations of the lattice localized near the interface. In the numerical simulations, the lattice size has been chosen large enough so as up the largest observation time edge effects at the boundaries of the integration domain are negligible (edge effects become important to possibly destroy interface states but only at long observation times, that increase with increasing system size, as discussed in \cite{r5}).
 In Fig.2 the lattice is initially excited in the single site $n=0$ at the interface: while for $h_1<0$, $h_2>0$ light remains trapped at the interface and $P(z)$ decreases [topological funneling, Fig.2(a)], for $0<h_1<h_2$   light is not trapped  anymore and flows toward the left medium being amplified [Fig.2(b)], according to theorem I. Figures 3 and 4 show the beam dynamics in the regime 
$0<h_1<h_2$ when the initial excitation of the lattice corresponds to a localized interface state [Eq.(6) with $\mu=(h_1+h_2)/2$] , perturbed by truncating either the body [panels (a)] or tails [panels (b)] of the exponentially-decaying eigenstate. In (a) the wave number $k$ is set to $k=\pi/2$, corresponding to a growth rate $g=2 \Delta \sinh \mu $ of the unperturbed interface state, while in (b) we set 
  $k=0$, corresponding to a vanishing growth rate $g=0$. Note that, when the cut is made on the tails of the interface state [Figs.3(b) and 4(b)], the initial excitation distribution $\psi_n(0)$ has a compact support
  and thus, according to theorem I, light trapping at the interface is not observed. Interestingly, the light beam remains trapped for a while, up to a time $t^*$, after which it flows in the left medium, resulting in a self-bending behavior [see e.g. Fig.3(b)]. A rough estimate of the trapping time $t^*$ is obtained by observing that, assuming a cut in the profile $\psi_n(0)$ at sites $|n| \geq N$, the front of the cut travels in the lattice with a speed $\sim 2 \Delta$, and thus it takes a time $t^* \sim N/2 \Delta$ to reach the body of the distribution and  to disrupt the eigenstate profile. When the cut is made in the body of the interface state distribution [Figs.3(a) and 4(a)], we perturb the eigenstate by a term which has a compact support, and hence theorem II applies. Accordingly, a beam self-healing effect and trapping are observed in Fig.3(a), where the growth rate of the interface state is larger than $g_0$, but not in Fig.4(a), where the perturbation grows yielding a characteristic drift of excitation in the bulk of the left medium \cite{r12}.\\
  \begin{figure}[htb]
\centerline{\includegraphics[width=8.7cm]{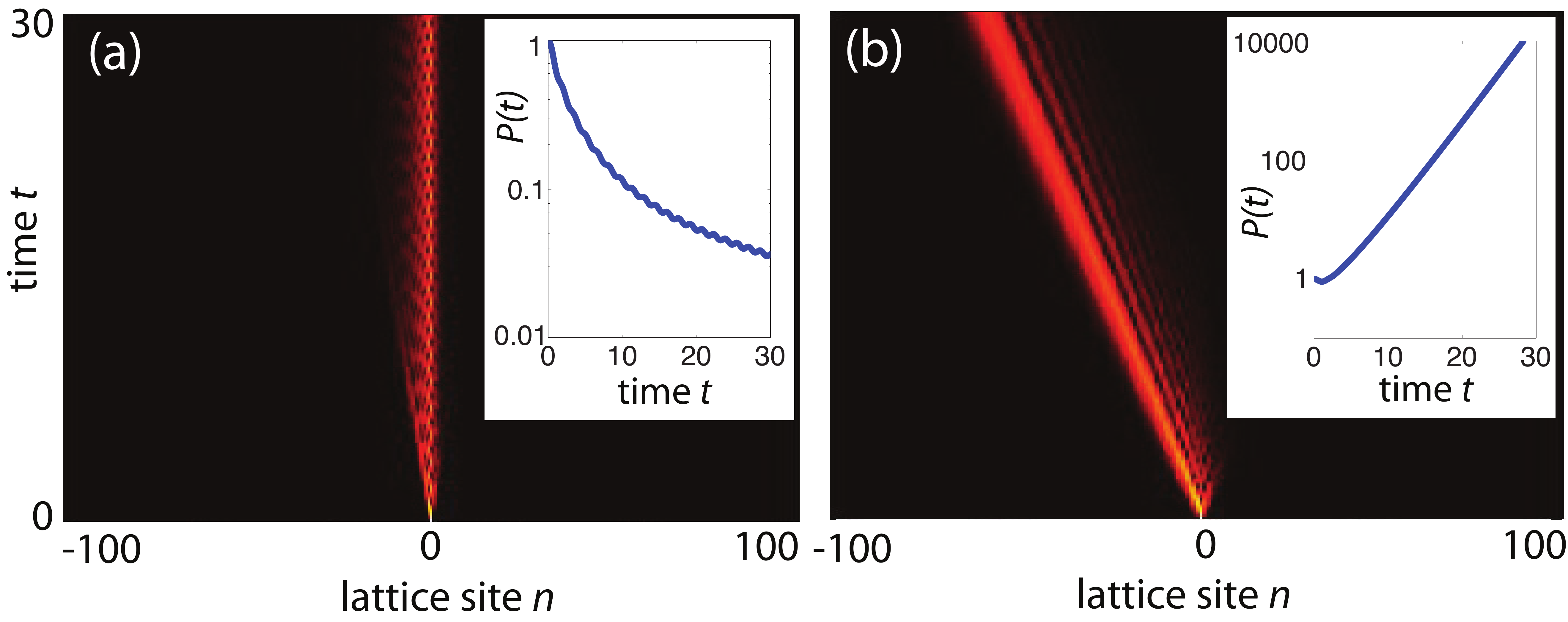}} \caption{ \small
(Color online) Light dynamics (temporal evolution of $|a_n(t)|$ on a pseudocolor map) in the Hatano-Nelson interface for single-site initial excitation $\psi_n(0)=\delta_{n,0}$  and for (a) $h_1=-0.1$, $h_2=0.3$ and (b)  $h_1=0.1$, $h_2=0.3$. The insets show the temporal behavior of the beam power $P(t)$, normalized to its input value, on a log scale. Time is normalized to $1/ \Delta$.}
\end{figure} 

 \begin{figure}[htb]
\centerline{\includegraphics[width=8.7cm]{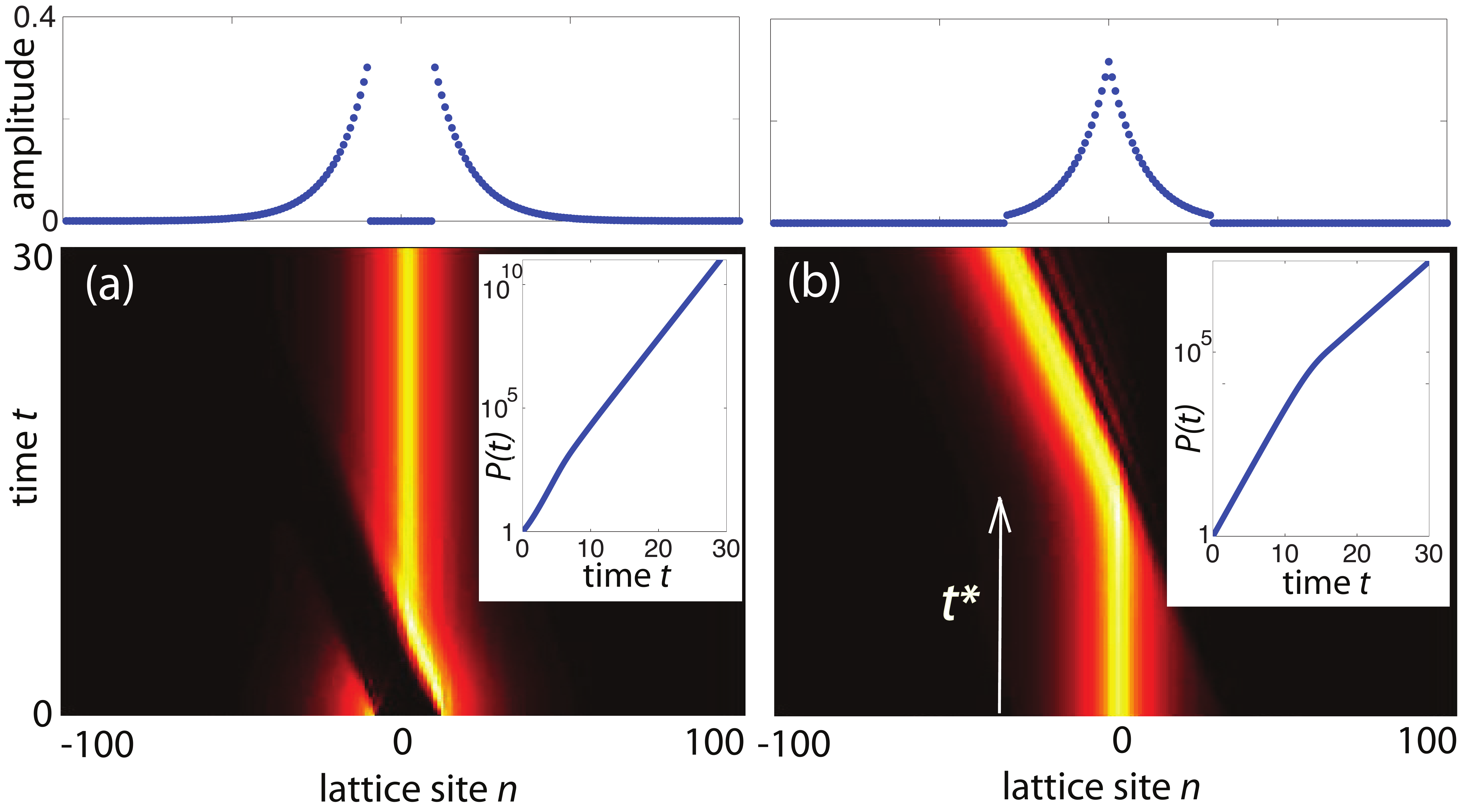}} \caption{ \small
(Color online) Light dynamics (temporal evolution of $|a_n(t)|$ on a pseudocolor map) in the Hatano-Nelson interface for $h_1=0.1$ and $h_2=0.3$. The lattice is initially excited in the interface eigenstate given by Eq.(6) with $\mu=(h_1+h_2)/2=0.2$ and $k=\pi/2$, either truncated at the body ($\psi_n(0)=0$ for $|n|<N=10$) or at the tails ($\psi_n(0)=0$ for $|n|>N=30$), as schematically shown in the upper panels of (a) and (b) where the behavior of $|\psi_n(0)|$ is depicted. Note that in (a) beam self-healing is observed, while in (b) light trapping is observed up to a time $t^*$, after which the beam flows in the left medium. The trapping time $t^*$ is roughly given by $t^* \sim N/ 2 \Delta=15/ \Delta$.}
\end{figure}

 \begin{figure}[htb]
\centerline{\includegraphics[width=8.7cm]{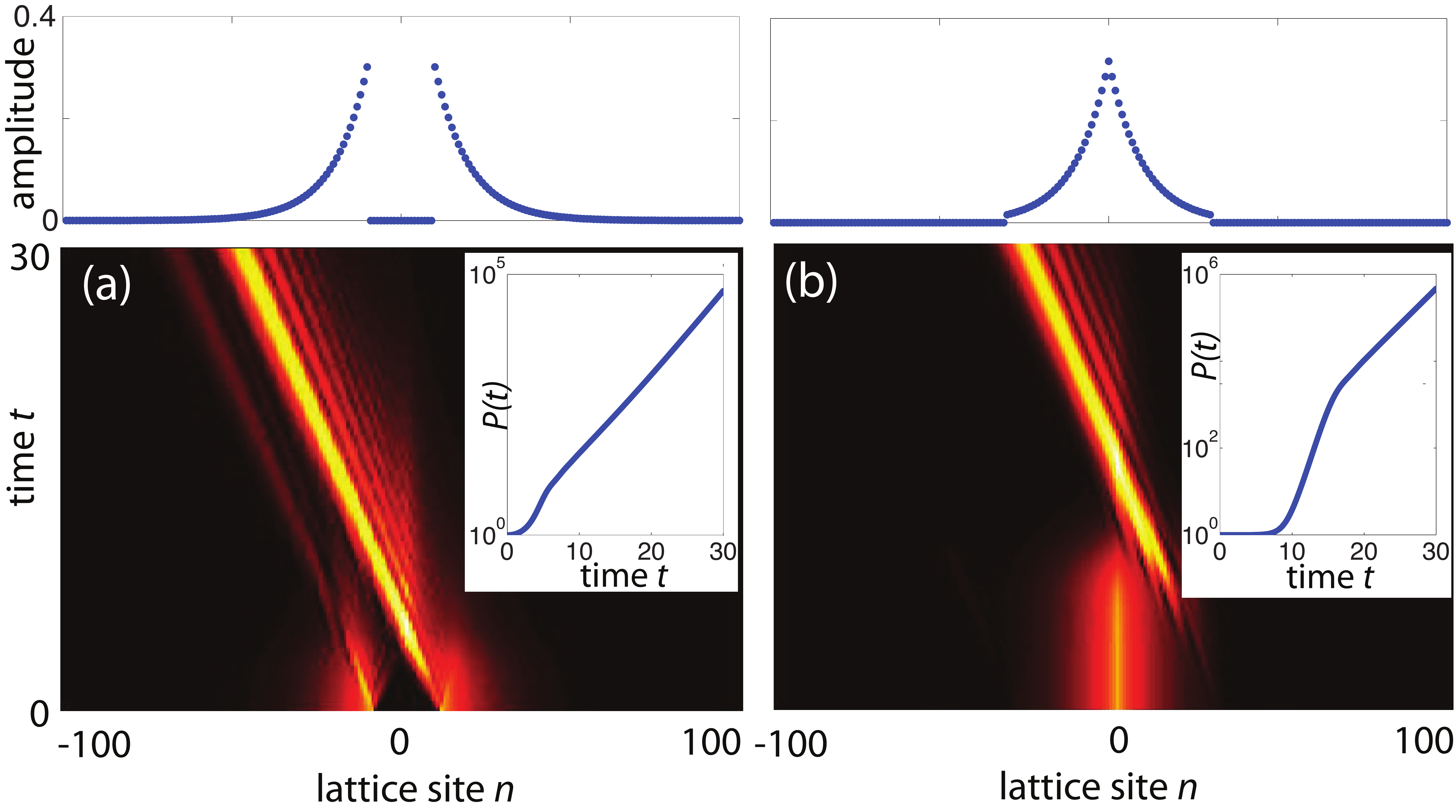}} \caption{ \small
(Color online) Same as Fig.3, but for  a wave number $k=0$.}
\end{figure} 

 {\em Photonic quantum walk at a NH interface.} The above analysis on bulk-edge correspondence and light trapping at a NH topological interface has been illustrated by considering the single-band Hatano-Nelson model, however the analysis could be extended to other NH topological models, including NH discrete-time photonic quantum walks \cite{r14,r16,r19}, and even beyond tight-binding models \cite{r30}. As an example, we consider a two-band model of photonic quantum walk realized in in coupled optical fiber loops \cite{r31}. Here, the imaginary gauge field $h$ is introduced by balanced gain/loss  in the two fiber loops \cite{r19}. Light dynamics  is governed by the discrete-time coupled equations \cite{r4,r19,r31}
 \begin{eqnarray}
 u^{(m+1)}_n & = & \left[   \cos \beta u^{(m)}_{n+1}+i \sin \beta v^{(m)}_{n+1}  \right]  \exp [h(n)] \\
 v^{(m+1)}_n & = & \left[   \cos \beta v^{(m)}_{n-1}+i \sin \beta u^{(m)}_{n-1}  \right] \exp [-h(n)]
 \end{eqnarray}
 where $u^{(m)}_n$, $v_n^{(m)}$ are the pulse amplitudes at lattice position $n$ and at discrete time step $m$ on the left and right moving paths, respectively, and $\beta$ is the coupling angle of the beam splitter ($\beta= \pi/4$ for a balanced 50/50 beam splitter). A NH topological interface is obtained by letting $h(n)=h_1$ for $n <0 $ and $h(n)=h_2$ for $n \geq 0$. The system sustains localized eigenstates at the interface of the form 
 $(u_n^{(m)}, v_n^{(m)})^T=(A_{\pm},B_{\pm}) \exp( \mu n +i kn-V(n)-i E_{\pm}m)$ with quasi energies given by
 \begin{equation}
 E_{\pm}= \pm {\rm acos} \left\{  \cos \beta \cos (k-i\mu) \right\}
 \end{equation}
 provided that $h_1 < \mu < h_2$. where $V(n)$ is defined by Eq.(5). As shown in the Supplemental document, a bulk-boundary correspondence can be established, i.e. at a complex quasi energy $E$ the interface states do exist provided that the corresponding  winding numbers $w_1(E)$ and $w_2(E)$ is the two media are distinct. In the $h_1<0, h_2>0$ case topological light funneling is observed \cite{r19}, while in the $0 \leq h_1< h_2$ case light trapping is prevented for any initial excitation of the system with compact support or with a localization higher than exponential. This behavior, which is analogous to the one of Fig.2 previously found for the Hatano-Nelson model, is illustrated in Fig.5.

 \begin{figure}[htb]
\centerline{\includegraphics[width=8.7cm]{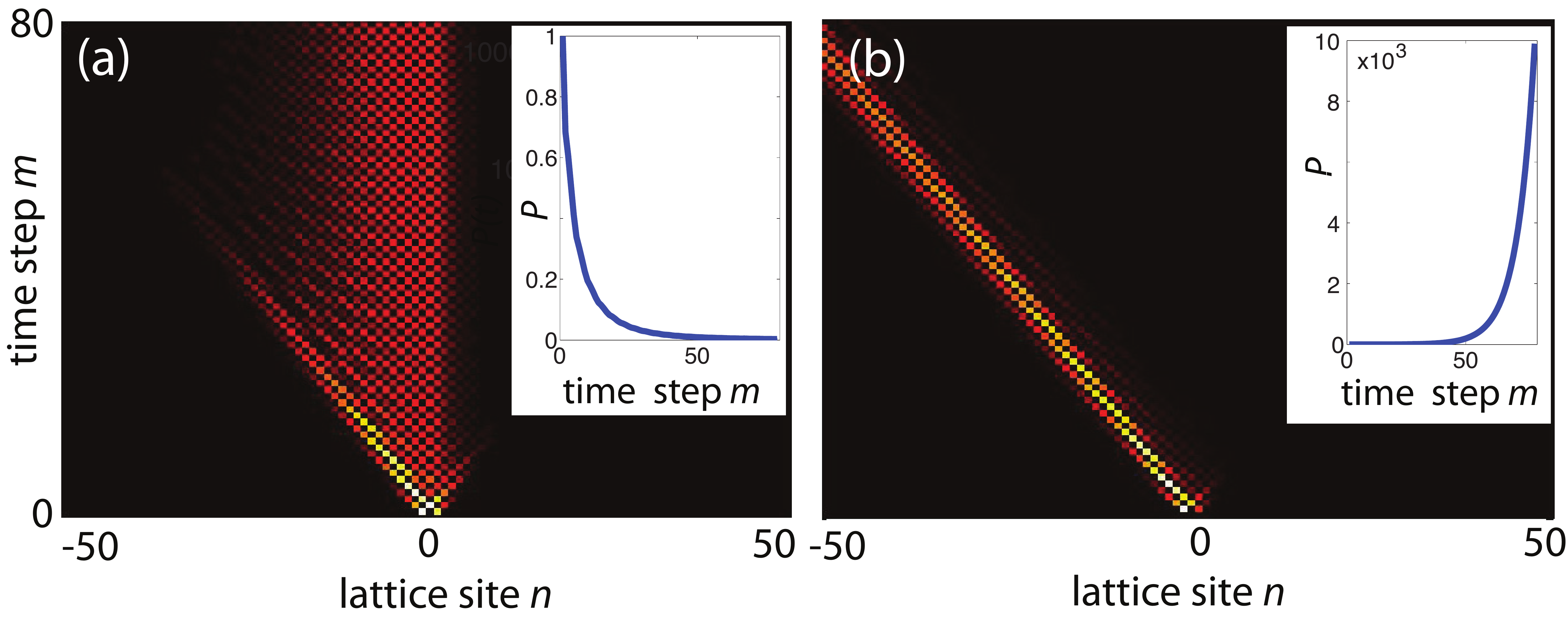}} \caption{ \small
(Color online) Light dynamics in a NH topological interface realized by fiber loops. The panels show on a pseudo color map the discrete-time evolution of $|u_n^{(m)}|^2+|v_n^{(m)}|^2$, normalized to the total optical power $P^{(m)}= \sum_n |u_n^{(m)}|^2+|v_n^{(m)}|^2$, in a lattice with balanced beam splitter $\beta= \pi/4$ for the initial excitation $u_n^{(0)}=v^{(0)}_n= \delta_{n,0}$ and for (a) $h_1=-0.1$, $h_2=0.3$, and (b) $h_1=0.1$, $h_2=0.3$. In (a) light is trapped at the interface, which behaves like a topological funnel, while in (b) it flows in the left medium. The insets show the temporal behavior of the total beam power $P^{(m)}$, normalized to its input value.}
\end{figure}   

{\em Conclusions.} The bulk-edge correspondence states that localized interface states should exist at the interface of two topological optical media with different topological numbers. This implies that for an arbitrary excitation of the system some light remains trapped at the interface, while some light spreads into the bulk. When turning to NH media, such a scenario is deeply modified owing to the appearance of the NH skin effect. In particular, an interface behaves as a topological funnel when the skin effect in the two media occurs in opposite directions: any light excitation fully flows toward the interface \cite{r19}. In this work we extended the bulk-edge correspondence to the NH realm and predicted that edge states should exist at a topological interface even when the skin effect occurs in the same direction. In spite of the existence of infinitely many localized states, in this case light can or cannot be trapped at the interface, depending on the initial excitation condition. In particular, for any spatially-localized excitation with a higher-than-exponential localization, light is never trapped and fully flows in the bulk. The present results provide major insights into the physics of NH topological systems and might be of potential interest to applications such as optical sensing, beam self-bending and beam shaping.

,

\par

\noindent
{\bf Disclosures}. The author declares no conflicts of interest.\\
\\
{\bf Acknowledgment}. The author acknowledges the Spanish
State Research Agency, through the Severo-Ochoa and Maria de
Maeztu Program for Centers and Units of Excellence in R\&D
(Grant No. MDM-2017-0711).\\
\\
{\bf Data Availability.} No data were generated or analyzed in the presented
research.\\
\\
{\bf Supplemental document}. See Supplement 1 for supporting content.

\newpage


\clearpage
 {\bf References with full titles}\\
 \\

 \noindent
 
1. X.-L. Qi, Y.-S. Wu, and S.-C. Zhang, General theorem relating the bulk topological number to edge states in two-dimensional insulators,
Phys. Rev. B {\bf 74}, 045125 (2006).

2. L. Lu, J.D. Joannopoulos, and M. Solja\v{c}ic , Topological Photonics, Nature Photon. {\bf 8}, 821 (2014).

3. T. Ozawa, H.M. Price, A. Amo, N. Goldman, M. Hafezi, L. Lu, M.C. Rechtsman, D. Schuster, J. Simon, O. Zilberberg, and I. Carusotto,
Topological Photonics, Rev. Mod. Phys. {\bf 91}, 015006 (2019).

4. S. Weidemann, M. Kremer, S. Longhi, and A. Szameit, Coexistence of dynamical delocalization and spectral localization through stochastic dissipation,
Nature Photon. {\bf 15}, 576 (2021).

5. Z. Gong, Y. Ashida, K. Kawabata, K. Takasan, S. Higashikawa, and M. Ueda, Topological Phases of Non-Hermitian Systems, Phys. Rev. X {\bf 8}, 031079 (2018).

6. B. Midya, H. Zhao, and L. Feng, Non-Hermitian photonics promises exceptional topology of light, 
Nature Commun. {\bf 9}, 2674 (2018).

7. K. Kawabata, K. Shiozaki, M. Ueda, and M. Sato, Symmetry and Topology in Non-Hermitian Physics, Phys. Rev. X 9, 041015 (2019).

8.  E.J. Bergholtz, J.C. Budich, and F.K. Kunst, Exceptional Topology in non-Hermitian Systems, Rev. Mod. Phys.{\bf  93}, 15005 (2021).

9. F.K. Kunst, E. Edvardsson, J.C. Budich, and E.J. Bergholtz, Biorthogonal Bulk-Boundary Correspondence in Non-Hermitian Systems, Phys. Rev. Lett. {\bf 121}, 026808 (2018).

10. S. Yao and Z. Wang, Edge States and Topological Invariants of Non-Hermitian Systems, Phys. Rev. Lett. {\bf 121}, 086803 (2018).

11. C.H. Lee and R. Thomale, Anatomy of skin modes and topology in non-Hermitian systems, Phys. Rev. B {\bf 99}, 201103(R) (2019).

12. S. Longhi, Probing non-Hermitian skin effect and non-Bloch phase transitions,
Phys. Rev. Research {\bf 1}, 023013 (2019).

13. F. Song, S. Yao, and Z. Wang, Non-Hermitian Topological Invariants in Real Space
Phys. Rev. Lett. {\bf 123}, 246801 (2019).

14.  T. Deng, K. Wang, G. Zhu, Z. Wang, W. Yi, and P. Xue,
Observation of non-Hermitian bulk-boundary correspondence in
quantum dynamics, Nat. Phys. {\bf 16}, 761 (2020).

15.  L Jin and Z Song, Bulk-boundary correspondence in a non-Hermitian system in one dimension with chiral inversion symmetry,
Phys. Rev. B {\bf 99}, 081103 (2019).

16. S. Longhi, Non-Bloch PT symmetry breaking in non-Hermitian photonic quantum walks,
Opt. Lett. {\bf 44}, 5804 (2019).

17. H. Zhao, X. Qiao, T. Wu, B. Midya, S. Longhi. and L. Feng,  Non-Hermitian topological light steering, Science {\bf 365}, 1163 (2019).

18. N. Okuma, K. Kawabata, K. Shiozaki, and M. Sato, Topological Origin of Non-Hermitian Skin Effects,
Phys. Rev. Lett. {\bf 124}, 086801 (2020).

19. S. Weidemann, M. Kremer, T. Helbig, T. Hofmann, A. Stegmaier, M. Greiter, R. Thomale, and A. Szameit, Topological funneling of light, Science {\bf 368}, 311 (2020).

20. Y. Song, W. Liu, L. Zheng, Y. Zhang, B. Wang, and P. Lu, 
Two-dimensional non-Hermitian Skin Effect in a Synthetic Photonic Lattice,
Phys. Rev. Applied {\bf 14}, 064076 (2020).

21. S. Longhi, Non-Bloch-Band Collapse and Chiral Zener Tunneling,
Phys. Rev. Lett. {\bf 124}, 066602 (2020).

22. K. Wang, A. Dutt, K. Y. Yang, C.C. Wojcik, J. Vuckovic, and S. Fan,
Generating arbitrary topological windings of a non-Hermitian band, Science {\bf 371}, 1240 (2021).

23. L. Xiao, T. Deng, K. Wang, Z. Wang, W. Yi, and P. Xue, Observation of Non-Bloch Parity-Time Symmetry and Exceptional Points,
Phys. Rev. Lett. {\bf 126}, 230402 (2021).

24. Z. Lin, S. Ke, X. Zhu, and X. Li,
Square-root non-Bloch topological insulators in
non-Hermitian ring resonators, Opt. Express {\bf 29}, 8462 (2021).

25. Z. Lin, L. Ding, S. Ke, and X. Li, 
Steering non-Hermitian skin modes by synthetic gauge fields in optical ring resonators, 
Opt. Lett. {\bf 46}, 3512 (2021).

26. H. C. Wu, L. Jin, and Z. Song, Topology of an anti-parity-time symmetric non-Hermitian Su-Schrieffer-Heeger model,
Phys. Rev. B {\bf 103}, 235110 (2021).
.
27. N. Hatano and D.R. Nelson, Localization Transitions in Non-Hermitian Quantum Mechanics, Phys. Rev. Lett. {\bf 77}, 570 (1996).

28. S. Longhi, D. Gatti, and G. Della Valle, Robust light transport in non-Hermitian photonic lattices, Sci. Rep. {\bf 5}, 13376 (2015).

29. S. Longhi, D. Gatti, and G. Della Valle, Non-Hermitian transparency and one-way transport in low-dimensional lattices by an imaginary gauge field, Phys. Rev. B {\bf 92}, 094204 (2015).

30. S. Longhi, Non-Hermitian skin effect beyond the tight-binding models, Phys. Rev. B {\bf 104}, 125109 (2021).

31. A. Schreiber, K. N. Cassemiro, V. Potocek, A. Gabris, P. J. Mosley, E. Andersson, I. Jex, and Ch. Silberhorn, Photons Walking the Line: A Quantum Walk with Adjustable Coin Operations,
Phys. Rev. Lett. {\bf 104}, 050502 (2010).

\end{document}